\documentclass[20pt,APS,balancelastpage,nofootinbib,twocolumn,fleqn,showpacs]{revtex4}%
\usepackage{amsfonts}
\usepackage{amsmath}
\usepackage{amssymb}
\usepackage[colorlinks, citecolor=blue,linkcolor=blue]{hyperref}
\usepackage{color}
\usepackage{float}
\usepackage{hyperref}
\usepackage{textcomp}
\usepackage{subfigure}
\usepackage[rightcaption]{sidecap}
\usepackage{graphicx}
\begin{document}
\title{Controlled Electromagnetically Induced Transparency and Fano Resonances in Hybrid BEC-Optomechanics}
\author{Kashif Ammar Yasir}
\email{kashif\_ammar@yahoo.com}
\author{Wu-Ming Liu}
\email{wliu@iphy.ac.cn}
\affiliation{Beijing National Laboratory for Condensed Matter Physics, Institute of Physics, Chinese Academy of Sciences, Beijing 100190, China.}

\begin{abstract}
We investigate the controllability of electromagnetically induced transparency (EIT) and Fano resonances in hybrid 
optomechanical system which is composed of cigar-shaped Bose-Einstein condensate (BEC) trapped inside high-finesse 
Fabry-P\'{e}rot cavity driven by a single mode optical field along the cavity axis and a transverse pump field. 
Here, transverse optical field is used to control the phenomenon of EIT in the 
output probe laser field. The output probe laser field can efficiently be amplified or attenuated depending on 
the strength of transverse optical field. Furthermore, we demonstrate the existence of Fano resonances in the output field 
spectra and discuss the controlled behavior of Fano resonances using transverse optical field. To observe this phenomena in laboratory, 
we suggest a certain set of experimental parameters.
\end{abstract}
\date{\today}
\maketitle

\section{Introduction}

Over a few years, Cavity-optomechanics, a rapidly developing area of research, has made a remarkable progress.
A stunning manifestation of optomechanical phenomena is in exploiting the mechanical effects of light to couple the optical 
degree of freedom with mechanical degree of freedom \cite{Kippenberg}. In this regard, a milestone was achieved by the demonstration of optomechanics when other physical objects, most 
notably cold atoms or Bose-Einstein condensate (BEC), were trapped inside cavity-optomechanics \cite{Esslinger,Ritsch2013}. 
Through several investigations, different aspects of BEC have made spectacular contributions in understanding the complex systems\cite{Qian,Vincent,spinorbit,Dong14}.
In optomechanical systems coupling is obtained by radiation pressure \cite{Braginsky1977,Mancini2002} and 
indirectly via quantum dot \cite{Tian2004} and ions \cite{Naik2006}. Optomechanics helps mechanical effects of light 
to cool the movable mirror to its quantum mechanical ground 
state \cite{CornnellNat2010,TeufelNat2011,CannNat2011,wmliu1,wmliu2} and provides a platform to study strong coupling 
effects in hybrid systems~\cite{GroeblacherNat2009,ToefelNat2011a,VerhagenNat2012}. 
Recent magnificent discussions and simulations on bistable behavior of BEC-optomechanical system
\cite{Meystre2010}, high fidelity state transfer \cite{YingPRL2012,SinghPRL2012}, entanglement in cavity-optomechanics
~\cite{chiara2011,Vitali2012,Vitali2014,Sete2014,Shi2010}, dynamical localization in field of cavity-optomechanics \cite{kashif1} and the coupled arrays 
of micro-cavities \cite{wmliu3} provide clear understanding for
cavity-optomechanics. 

Recently, electromagnetically induced transparency (EIT), a phenomenon of direct manifestation of quantum 
coherence \cite{Scully,SafaviNaeini2011}, has been extensively investigated and has provided a lot of remarkable applications \cite{Harris1,Harris2}.
In atomic system, the EIT occurs due to quantum interference effects induced by coherently driving atomic wavepacket with an 
external pump laser field \cite{Kash}. The EIT effect has been theoretically investigated in optomechanical system \cite{agarwal2010,Asjad2013} 
and later experimentally verified in both optical \cite{Chang,Stefan2010} and microwave \cite{Massel2012} domains.
The Fano resonances \cite{Fano} have played an important role in understanding the photo-electron spectra \cite{Bransden2003} in atomic
physics and have made a magnificent contribution in the latest field of plasmonics \cite{Gallinet}. Fano resonances have
also been investigated in hybrid cavity-optomechanics by using different configuration \cite{agarwal2013,Jadi2015}.

In this paper, we investigate the controlled behavior of electromagnetically induced transparency (EIT) and Fano Resonances in hybrid 
BEC-optomechanical system which is composed of cigar-shaped Bose-Einstein condensate (BEC) trapped inside high-finesse 
Fabry-P\'{e}rot cavity, with one fixed mirror and one moving-end mirror with maximum amplitude $q_0$, driven by 
a single mode optical field $\eta$ along the cavity axis and a transverse pump field $\eta_{\perp}$. 
Transverse optical field is used to control the phenomenon of electromagnetically induced transparency (EIT) in the 
output probe laser field. By observing output probe field spectra, we show that the probe laser field can efficiently be 
amplified or attenuated depending on the strength of transverse optical field $\eta_{\perp}$. Furthermore, we demonstrate the existence 
of Fano resonances by solving output field spectra and discuss the controlled behavior of Fano resonances using transverse 
optical field. To observe this phenomena in laboratory, we have suggested a certain set of experimental parameters.

In this paper, the mathematical model of the hybrid BEC-optomechanical system is presented in section~\ref{sec1}. 
In section~\ref{sec2}, we derive the Langevin equation to introduce the dissipation effects in the intra-cavity field, 
damping associated with moving-end mirror and depletion of BEC in the system via standard quantum noise operators. 
In section~\ref{sec3}, we discuss EIT in output probe field of the hybrid cavity-optomechanics and explain controlled 
behavior of EIT with transverse optical field. In section~\ref{sec4}, 
we demonstrate the existence of fano resonances and describe their controllability by using transverse optical field. 
The results are summarized in section-\ref{sec5}.

\section {BEC-Optomechanics}\label{sec1}

\begin{figure}
\includegraphics[width=8.5cm]{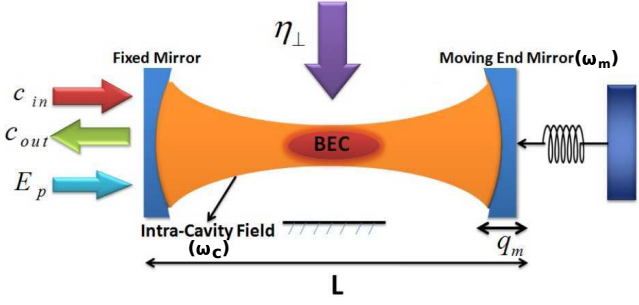}
\caption{(Color online) Cigar-shaped Bose-Einstein condensate (BEC) is trapped inside a Fabry-P\'{e}rot cavity of length $L$
with a fixed mirror and a moving-end mirror, with frequency $\omega_{m}$ and having maximum amplitude of $q_0$, driven by a single mode 
optical field with frequency $\omega_{p}$. Intracavity field, with frequency $\omega_{c}$, interacts with BEC and due 
to photon recoil, the momentum side modes are generated in the matter wave. To observe electromagnetically induced 
transparency (EIT) in output field spectra, we use external probe field with frequency $\omega_{pp}$. 
$E_{p}$ corresponds to the power of external probe field. 
We use another transverse field, with intensity $\eta_\perp$, to control the 
output spectral dynamics of optomechanical system.}
\label{fig1}
\end{figure}

We consider a Fabry-P\'{e}rot cavity of length $L$ with a fixed mirror and a moving-end mirror driven by a single mode 
optical field with frequency $\omega_{p}$, as shown in Fig. \ref{fig1}. Cigar-Shaped BEC, with N-two level atoms, is trapped inside optical cavity
\cite{Esslinger,esteve,Brennecke}. Moreover, optomechanical system consists of a transverse optical field, with intensity $\eta_\perp$ and frequency $\omega_{\perp}$, which interacts 
perpendicularly with BEC \cite{chiara2010,Kashif2015,Zhang2009,Yang2011,zhang2013}. Counter propagating field inside the cavity forms a one-dimensional optical lattice.
The moving-end mirror has harmonic oscillations with frequency $\omega_{m}$ having a maximum amplitude $q_0$ and 
exhibits Brownian motion in the absence of coupling with radiation pressure.

The complete Hamiltonian of the system consists of three parts,
\begin{equation}
\hat{H}=\hat{H}_{m}+\hat{H}_{a}+\hat{H}_{T}\label{1},
\end{equation}
where, $\hat{H}_{m}$ is related to the intra-cavity field and its coupling to the moving-end mirror, $\hat{H_{a}}$ 
describes the BEC and its coupling with intra-cavity field while, $\hat{H}_{T }$ accounts for noises and 
damping associated with the system. The Hamiltonian $\hat{H}_{m}$ is given as \cite{LawPRA1995},
\begin{eqnarray}
\hat{H}_{m} &=& \hbar\triangle_{c}\hat{c}^{\dag}\hat{c}+\frac{\hbar\omega_{m}}{2}(\hat{p}^{2}+
\hat{q}^{2})-g_{m}\hbar\hat{c}^{\dag}\hat{c}\hat{q}-i\hbar\eta(\hat{c}-\hat{c}^{\dag}) \nonumber \\ &-&
i\hbar E_{p}(\hat{c}e^{i\Delta_{p}}-\hat{c}^{\dag}e^{-i\Delta_{p}}),
\end{eqnarray}
where, first term corresponds to the energy of the intra-cavity field, $\Delta_{c}=\omega_{c}-\omega_{p}$ is detuning, 
$\omega_{c}$ is intra-cavity field frequency and $\hat{c}^{\dag}$ ($\hat{c}$) are creation (annihilation) operators 
for intra-cavity field interacting with mirror and condensates with commutation relation  
$[\hat{c},\hat{c}^{\dag}]=1$. Second term accounts for the energy of moving-end mirror. Here $\hat{q}$ and $\hat{p}$ are 
representing dimensionless position and momentum operators respectively, for moving-end mirror with commutation relation $[\hat{q},\hat{p}]=i$,
which reveals the value of the scaled Planck's constant, $\hbar=1$. 
As intra-cavity field is coupled with moving-end mirror through radiation pressure so third term
represents coupled energy 
of moving-end mirror with field. Here $g_{m}=\sqrt{2}(\omega_{c}/L)x_{0}$ is the coupling strength and 
$x_{0}=\sqrt{\hbar/2m\omega_{m}}$, is zero point motion of mechanical mirror having mass $m$. 
Second last term gives relation between the intra-cavity field and the external pump field $\vert\eta\vert=\sqrt{P_{in}\kappa/\hbar\omega_{p}}$, 
where, $P$ is the input laser power and $\kappa$ is cavity decay rate associated with out-going modes. Last term
is associated with external probe field and $E_p$ is related to the power of the probe field as 
$\vert E_{p}\vert=\sqrt{P_{p}\kappa/\hbar\omega_{p}}$. $\Delta_{p}=\omega_{p}-\omega_{l}$ is the detuning of external
probe laser field with external pump field.

The Hamiltonian for BEC and its coupling with intra-cavity field ($H_a$), in strong detuning regime and in the rotating frame at external
field frequency is derived by considering quantized motion of atoms along with the cavity axis. We assume that BEC is dilute enough and many body 
interaction effects are ignored\cite{Zhang2009,Yang2011}. Here,
\begin{eqnarray}
\hat{H}_a &=& \int\hat{\psi}^{\dag}(x)[-\frac{\hslash d^{2}}{2m_{a}dx^{2}}+
\hslash U_{0}\hat{c}^{\dag}\hat{c}\cos^{2}kx \nonumber \\ &+&
\hbar\eta_{\perp} cos(kx)(\hat{c}^{\dag}+\hat{c})]\hat{\psi}(x)dx,\label{BEC}
\end{eqnarray}
$\hat{\psi}(\hat{\psi}^{\dag})$ is bosonic annihilation (creation) operator and 
$U_{0}=g^2_{0}/\Delta_{a}$, $g_{0}$ is the vacuum Rabi frequency, $\Delta_{a}$ is far-off detuning between 
field frequency and atomic transition frequency. [Note: in equation \ref{BEC}, we have not considered for now the effects of 
harmonic trap causing the confinement of atoms inside the cavity.] Furthermore, $m_a$ is mass of an atom, $\omega_{0}$ is atomic 
transition frequency and $k=\omega_p/c$ is the wave number. $\eta_\perp=g_{0}\Omega_{p}/\Delta_a$ is the coupling of BEC with transverse
field and represents maximum scattering and $\Omega_p$ is the Rabi frequency of the transverse pump field.
Due to field interaction with BEC, photon recoil takes place that generates symmetric momentum $\pm2{l}\hbar k$ side modes, where, ${l}$ is an integer. 
In the weak field approximation, we consider low photon coupling. Therefore, only lowest order perturbation of the wave function 
will survive and higher order perturbation will be ignored.  
So, $\hat{\psi}$ is defined depending upon these $0^{th}$, $1^{st}$ and $2^{nd}$ modes~\cite{Meystre2010} as,
\begin{equation}
 \hat{\psi}(x)=\frac{1}{\sqrt{L}}[\hat{b_{0}}+\sqrt{2}cos(kx)\hat{b_{1}}+\sqrt{2}cos(2kx)\hat{b_{2}}],
\end{equation}
here, $\hat{b_{0}}$,$\hat{b_{1}}$ and $\hat{b_{2}}$ are annihilation operators for $0^{th}$ 
, $1^{st}$ and $2^{nd}$ modes respectively. By using $\hat{\psi}(x)$ defined above in Hamiltonian $H_a$, we write the Hamiltonian 
governing the field-condensate interaction as,
\begin{eqnarray}
  \hat{H}_{a} &=& \omega_{r}\hat{b_{1}}^{\dag}\hat{b_{1}}+ 4\omega_{r}\hat{b_{2}}^{\dag}\hat{b_{2}}+
 \frac{\hbar U_{0}}{4}\hat{c}^{\dag}\hat{c}[\sqrt{2}(\hat{b_{0}}^{\dag}\hat{b_{2}}+\hat{b_{2}}^{\dag}\hat{b_{0}}) \nonumber \\
 &+& 2N+\hat{b_{1}}^{\dag}\hat{b_{1}}]+\frac{\hbar \eta_{\perp}}{2}(\hat{c}^{\dag}+\hat{c})[\sqrt{2}(\hat{b_{0}}^{\dag}\hat{b_{1}} \nonumber \\
 &+& \hat{b_{1}}^{\dag}\hat{b_{0}})+(\hat{b_{1}}^{\dag}\hat{b_{2}}+\hat{b_{2}}^{\dag}\hat{b_{1}})].
\end{eqnarray}
The sum of particles in all momentum side modes is, 
$\hat{b_{0}}^{\dag}\hat{b_{0}}+\hat{b_{1}}^{\dag}\hat{b_{1}}+\hat{b_{2}}^{\dag}\hat{b_{2}}=N$, where, $N$ is the total number 
of bosonic particles. As population in $0^{th}$ mode is much larger than the population in $1^{st}$ and $2^{nd}$ order side modes, 
therefore, we can comparatively ignore the population in $1^{st}$ and $2^{nd}$ order side modes and can write 
$\hat{b_{0}}^{\dag}\hat{b_{0}}\simeq N$ or $\hat{b_{0}}$ and $\hat{b_{0}}^{\dag}\rightarrow \sqrt{N}$. This is possible 
when side modes are weak enough to be ignored. Moreover, for Bogoliubov mode expansion, we consider small interaction of 
intra-cavity field with BEC so, that atomic mode can perform motion analog to the moving-end mirror of the system. 
Under these assumptions, we recover the cavity-optomechanics-like Hamiltonian discussed in 
Ref.\cite{Esslinger} and \cite{Meystre2010} and given as,
\begin{equation}\label{Ha}
\hat{H}_{a}=\frac{\hbar U_{0}N}{2}\hat{c}^{\dag}\hat{c}+\frac{\hbar\Omega}{2}(\hat{P}^{2}
+\hat{Q}^{2})+g_{a}\hbar\hat{c}^{\dag}\hat{c}\hat{Q}+\hbar\eta_{eff}\hat{Q}(\hat{c}^{\dag}+\hat{c}).
\end{equation}
First term accommodates the potential energy for condensate in intra-cavity field. We assume large atom-field 
detuning $\Delta_{a}$, so that, the excited atomic levels can be adiabatically eliminated. Second term describes the motion 
atomic momentum side modes exited by radiation pressure. It can be observed that the atomic side modes are analogous of a mirror whose 
motion is driven by radiation pressure.
$\hat{P}=\frac{i}{\sqrt{2}}(\hat{b}_2-\hat{b}^{\dag}_2)$ and 
$\hat{Q}=\frac{1}{\sqrt{2}}(\hat{b}_2+\hat{b}^{\dag}_2)$ are dimensionless momentum and position operators for 
such atomic mirror with canonical relation, $[\hat{Q},\hat{P}]=i$ and $\Omega=4\omega_{r}=2\hbar k^{2}/m_{a}$, is recoil 
frequency of an atom. Third term in equ.(\ref{Ha}) describes coupled energy of field and condensate 
with coupling strength $g_{a}=\frac{\omega_{c}}{L}\sqrt{\hbar/m_{bec}4\omega_{r}}$, where, 
$m_{bec}=\hslash\omega_{c}^{2}/(L^{2}NU^2_{0}\omega_{r})$ is the side mode mass of condensate. The last term accounts
for the coupling of BEC with transverse field and $\eta_{eff}=\sqrt{n}\eta_{\perp}$ is transverse coupling strength. 
From equ.\ref{Ha}, it can be noted that in the absence of transverse optical field, when there is no excitation for $cos(kx)$, 
we recover the same expression for atomic mode of the cavity-optomechanics as in Ref.\cite{Esslinger} and \cite{Meystre2010}.

\section {Langevin Equations}\label{sec2}

The Hamiltonian $\hat{H}_{T}$ describes the effects of dissipation in the intra-cavity field, damping of 
moving-end mirror and depletion of BEC in the system via standard quantum noise operators ~\cite{Noise1991}.
The total Hamiltonian $H$ leads to develop coupled quantum 
Langevin equations for optical, mechanical (moving-end mirror) and atomic (BEC) degrees of freedom.
\begin{eqnarray}\label{2}
\frac{d\hat{c}}{dt}&=&\dot{\hat c}=(i\Delta_{0}+ig_{m}
\hat{q}-ig_{a} \hat Q-\kappa)\hat{c}+i\eta_{eff}\hat{Q} \nonumber \\ &+&
\eta + E_{p}e^{-i\Delta_{p}t}+\sqrt{2\kappa} c_{in},\label{2a}\\
\frac{d\hat{p}}{dt}&=&\dot{\hat p}=-\omega_{m}\hat{q}+g_{m}\hat{c}^{\dag}\hat{c}
-\gamma_{m}\hat{p}+\hat{f}_{B},\label{2b}  \\
\frac{d\hat{q}}{dt}&=&\dot{\hat q}=\omega_{m}\hat{p},\label{2c}  \\
\frac{d\hat{P}}{dt}&=&\dot{\hat P}=-4\omega_{r}\hat{Q}-g_{a}\hat{c}^{\dag}\hat{c}
-\gamma_{r}\hat{P}+\hat{f}_{1m},\label{2d} \\
\frac{d\hat{Q}}{dt}&=&\dot{\hat Q}=4\omega_{r}\hat{P}-\gamma_{r}\hat{Q}+\hat{f}_{2m}.\label{2e}
\end{eqnarray}
$\tilde{\Delta}=\Delta _{c}-NU_{0}/2$ is the effective detuning of the system and $\hat{a}_{\mathrm{in}}$ is Markovian input 
noise associated with intra-cavity field. The term $\gamma _{m}$ describes mechanical energy decay rate of the moving-end mirror and 
$\hat{f}_{B}$ is Brownian noise operator associated with the motion of moving-end mirror~\cite{Pater06}. 
The term $\gamma_{r}$ represents damping of 
BEC due to harmonic trapping potential which affects momentum side modes while, $\hat{f}_{1m}$ and $\hat{f}_{2m}$ 
are the associated noise operators assumed to be Markovian. We consider positions and momenta as classical variables. To write the steady state values of 
of the operator, we assume optical field decay at its fastest rate so that the time derivative can be set
to zero in equation (\ref{2a}). The static solutions are given as,
\begin{eqnarray}
 c_{s}&=&\frac{\eta+i\eta_{eff}Q}{\kappa +i(\Delta_{a}-g_{m} q +g_{a} Q)},\\
 q_{s}&=&\frac{g_{m} c^{\dag}c}{\omega_{m}},\\
 p_{s}&=& 0,\\
 Q_{s}&=&\frac{-g_{a}c^{\dag}c}{4\omega_{r}[1-\gamma_{r}/4\omega_{r}]},\\
 P_{s}&=&\frac{\gamma_{r}}{4\omega_{r}}Q_{s}
\end{eqnarray}\label{equ2}
where $c_{s}$, $q_{s}$ and $Q_{s}$ represent the steady-state solution of intra-cavity field, the mechanical mirror 
displacement, and the position of the BEC mode, respectively. To observe output field spectra, we deal with the mean response of 
the system to the probe field in the presence of the coupling field and transverse field. First, we linearized quantum 
Langevin equations by inserting ansatz $\hat{c(t)}=c_{s}+\delta c(t)$, $\hat{q(t)}=q_{s}+\delta q(t)$, 
$\hat{p(t)}=p_{s}+\delta p(t)$, $\hat{Q(t)}=Q_{s}+\delta Q(t)$ and $\hat{P(t)}=P_{s}+\delta P(t)$ in Langevin equations and
taking care of only first-order terms in fluctuating operators $\delta c(t)$, $\delta q(t)$, $\delta p(t)$, $\delta Q(t)$ 
and $\delta P(t)$. The linearized quantum Langevin equations are obtained as,
\begin{eqnarray}
 \delta\dot{c}(t) &=& -(\kappa+i\Delta)\delta c(t)+G_{m}\delta q(t)-G_{a}\delta Q(t)\nonumber \\ 
 &+&i\eta_{eff}\delta Q(t)+E_{p}e^{-i\Delta_{p}t}+\sqrt{2\kappa} c_{in},\\
 \delta\dot{c}^{\dag}(t) &=& -(\kappa-i\Delta)\delta c(t)+G_{m}\delta q(t)-G_{a}\delta Q(t)\nonumber \\ 
 &-&i\eta_{eff}\delta Q(t)+E_{p}e^{i\Delta_{p}t}+\sqrt{2\kappa} c_{in},\\
 \delta\dot{q}(t) &=& \omega_{m}\delta p(t),\\
 \delta\dot{p}(t) &=& -\omega_{m}\delta q(t)+G_{m}(\delta c(t)+\delta c^{\dag}(t))\nonumber \\ 
 &-&\gamma_{m}\delta p(t)+\hat{f}_{B},\\
 \delta\dot{Q}(t) &=& \omega_{r}\delta P(t)+\hat{f}_{2m},\\
 \delta\dot{P}(t) &=& -\omega_{m}\delta Q(t)+G_{a}(\delta c(t)+\delta c^{\dag}(t))\nonumber \\ 
 &-&\gamma_{r}\delta P(t)+\hat{f}_{1m},
 \end{eqnarray}
where, $\Delta=\Delta_{0}-g_{m}q+g_{a}Q$ is the effective detuning of the system and $G_{m}=g_{m}|c_s|$, $G_{a}=g_{a}|c_s|$ 
are the effective coupling of optical field with the moving-end mirror and the condensate
mode, respectively. To solve mean value equation of the system, we write expectation value of operators in form 
$O=\Sigma^{n=\infty}_{n=-\infty}e^{-n\omega t}O_{n}$, here $O$ is a generic operator. $<\delta c(t)>$, $<\delta q(t)>$, $<\delta p(t)>$, $<\delta Q(t)>$ 
and $<\delta P(t)>$ are the expectation values corresponding to fluctuating operators $\delta c(t)$, $\delta q(t)$, $\delta p(t)$, $\delta Q(t)$ 
and $\delta P(t)$, respectively.

To solve linearized quantum Langevin equations, we assume that the coupling of external pump field $\eta$ is much stronger than
the coupling of external probe field $E_p$. Under this assumption, the solutions of linearized Langevin equations can be
approximated to the first order external probe field by ignoring all higher order terms of $E_p$. So, the solution for intra-cavity
probe field is now given as,
\begin{eqnarray}
 \tilde{c}_{+} &=& \frac{E_{p}}{X(\Delta_{p})Y(\Delta_{p})}[Y(\Delta_{p})+(\kappa+i(\Delta+\Delta_{p}) \nonumber \\ 
 &+&X(\Delta_{p}))(\kappa-i(\Delta-\Delta_{p})+X(\Delta_{p}))]\label{11},\\ 
 \tilde{c}_{-} &=& E_{p}[\frac{\kappa-i(\Delta+\Delta_{p})+X^{*}(\Delta_{p})}{Y^{*}(\Delta_{p})}], \label{12}
\end{eqnarray}
where,
\begin{eqnarray}
 X(\Delta_{p}) &=& -\frac{G^{2}_{a}\omega_{r}+iG_{a}\eta_{eff}\omega_{r}}{\omega_{r}^{2}-\Delta^{2}_{p}+i\gamma_{r}\Delta_{p}}\nonumber \\ 
 &-&\frac{G^{2}_{m}\omega_{m}}{\omega_{m}^{2}-\Delta^{2}_{p}+i\gamma_{m}\Delta_{p}},\\
 Y(\Delta_{p}) &=& -\Delta^{2}-\kappa^{2}-2i\kappa\Delta_{p}+\Delta_{p}^{2} \nonumber \\ 
 &+&\frac{2\omega_{r}G_{a}^{2}(\kappa+i\Delta_{p})-2\omega_{r}\Delta\eta_{eff}G_{a}}{i\gamma_{r}\Delta_{p}-\Delta_{p}^{2}+\omega_{r}^{2}} \nonumber \\ 
 &+&\frac{2\kappa\omega_{m}G_{m}^{2}}{i\gamma_{m}\Delta_{p}-\Delta_{p}^{2}+\omega_{m}^{2}}.
 \end{eqnarray}
 
Eq.\ref{11} and Eq.\ref{12} clearly describe the dependence of output field spectra on the coupling of different 
degrees of freedom. In particular, we can observe the rule of transverse optical field coupling with BEC mode in output field.
Furthermore, to investigate the EIT-like behavior, we write output field spectra by using input-output relation 
$c_{out}=\sqrt{2\kappa}c-c_{in}$ \cite{Noise1991}, where $c_{in}$ and $c_{out}$ represent input and output field, respectively.
Moreover, we ignore quantum noises associated with $c_{out}$ and $c_{in}$ as discussed earlier. The out-going optical field can be expressed as,
\begin{eqnarray}
 <c_{out}> &=& c_{0}+c_{+}E_{p}e^{-i\Delta_{p}t}+c_{-}E_{p}^{*}e^{i\Delta_{p}t}
\end{eqnarray}

By using above relation and input-output field relation, we describe the components of output field spectra at 
probe field frequency and Stokes frequency as, 
\begin{eqnarray}
 c_{+} &=& \sqrt{2\kappa}\tilde{c}_{+}+1,\nonumber \\
 c_{-} &=& \sqrt{2\kappa}\tilde{c}_{-},
\end{eqnarray}
respectively. In the absence of optical coupling with 
moving-end mirror and BEC mode, the output field spectra at probe frequency and stokes frequency is given as,
\begin{eqnarray}
 c_{-} &=& \frac{2\kappa}{\kappa+i(\Delta-\Delta_{p})},\\
 c_{+} &=& 0. 
\end{eqnarray}

\section{Controllable EIT in Output Field}\label{sec3}
\begin{figure}[h]
\includegraphics[width=7.5cm]{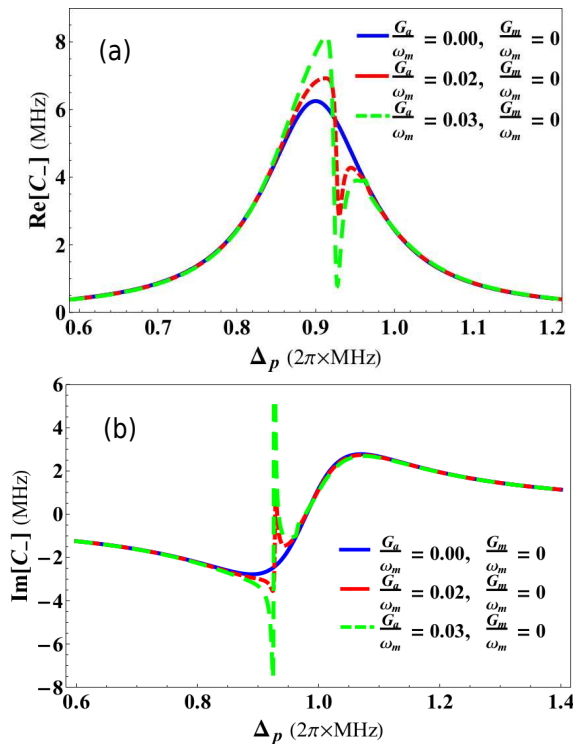}
\caption{(Color online) The absorption and dispersion quadratures of the output field spectra $C_{-}$ are demonstrated as a 
function of probe detuning $\Delta_{p}/\omega_{m}$ and in the absence of of transverse optical field coupling $\eta_{eff}/\kappa=0$.
(a) describe the behavior of single-EIT in real quadrature of output probe field with various coupling strengths $G_{a}/\omega_{m}=0$, 
$G_{m}/\omega_{m}=0$ (blue curve), $G_{a}/\omega_{m}=0.02$, $G_{m}/\omega_{m}=0$ (red curve), and $G_{a}/\omega_{m}=0.03$, $G_{m}/\omega_{m}=0$ (green curve).
(b) contains similar behavior in imaginary quadrature of output probe field with different couplings $G_{a}/\omega_{m}=0$, 
$G_{m}/\omega_{m}=0$ (blue curve), $G_{a}/\omega_{m}=0.02$, $G_{m}/\omega_{m}=0$ (red curve), and $G_{a}/\omega_{m}=0.03$, $G_{m}/\omega_{m}=0$ (green curve).
The effective detuning of the system is $\Delta=0.52\times2\pi MHz$ and the frequency of moving-end mirror is considered as, $\omega_{m}=1.02\times2\pi MHz$. 
The atomic mode and mechanical mode damping are $\gamma_r=0.21\times2\pi kHz$ and $\gamma_m=1.1\times2\pi kHz$, respectively. The cavity 
optical decay is $\kappa=1.3\times2\pi kHz$.}
\label{fig2a}
\end{figure}
The hybrid BEC-optomechanical system shown in Fig.\ref{fig1} is simultaneously driven by external pump field with frequency $\omega_{p}$ 
and probe field with frequency $\omega_{pp}$, generating radiation pressure force which oscillates at frequency difference 
$\Delta_{p}=\omega_{pp}-\omega_{p}$. When this resultant force oscillates with frequency close to the frequency of mechanical mode (moving-end mirror) $\omega_{m}$ or atomic mode (BEC) 
$4\omega_{r}$, it gives rise to the Stokes and anti-Stokes scatterings of light from the strong intra-cavity standing field. 
But the Stokes scattering is strongly suppressed because conventionally, optomechanical systems are operated in resolved-sideband regime $\kappa<<\omega_{m}$, 
which is off-resonant with Stokes scattering and so only anti-Stokes scattering survives inside the cavity. Therefore, due to the 
presence of probe field and pump field inside the system, electromagnetically induced transparency (EIT) like behavior appears 
in the output field spectra. 

To make this study of tunable EIT and Fano resonances in hybrid BEC-Optomechanics experimentally feasible, we choose a regime 
of particular parameters very close to the recent experimental studies \cite{Esslinger,Kippenberg,Brennecke}. In addition, we 
consider the parameters such that the system remain in stable regime as discuss in Ref. \cite{Yang2011,Kashif2015}.
We consider $N=2.3\times10^4$ $^{87}Rb$ atoms trapped inside Fabry-P\'{e}rot cavity with length $L=1.25\times10^{-4}$,
driven by single mode external field with power $P_{in}=0.0164mW$, frequency $\omega_{p}=3.8\times2\pi\times10^{14}Hz$ and
wavelength $\lambda_{p}=780nm$. The intra-cavity optical mode oscillates with frequency
$\omega_{c}=15.3\times2\pi\times10^{14}Hz$, with intra-cavity decay rate $\kappa=1.3\times2\pi kHz$. The vacuum Rabi 
frequency of the system is considered as, $U_{0}=3.1\times2\pi MHz$. Intra-cavity field produces
recoil of $\omega_{r}=3.8\times2\pi kHz$ in atomic mode trapped inside cavity with damping rate $\gamma_r=0.21\times2\pi kHz$. The moving-end mirror of cavity should be a 
perfect reflector and oscillates with frequency $\omega_{m}=1.02\times2\pi MHz$ with damping $\gamma_m=1.1\times2\pi kHz$. 
From given parameters, one can observe that the system is in resolved-sideband regime because $\omega_m>>\kappa$, this 
condition is also referred to good-cavity limit.

The real ($Re[C_-]$) and imaginary ($Im[C_-]$) quadratures of out-going probe field as a function of probe-cavity detuning are 
discussed in Fig\ref{fig2a}, in the absence of transverse field coupling $\eta_{eff}/\kappa=0$. Here, $Re[C_-]$ and $Im[C_-]$ 
accounts for in-phase and out-phase, respectively, quadratures of output field spectra and also referred to the absorption and 
dispersion behavior of out-going optical mode. Fig.\ref{fig2a}(a) and Fig.\ref{fig2a}(b) demonstrate the single-EIT behavior of 
such absorption and dispersion quadratures, respectively, of output field spectra for different coupling strengths $G_{a}/\omega_{m}=0$, 
$G_{m}/\omega_{m}=0$ (blue curve), $G_{a}/\omega_{m}=0.02$, $G_{m}/\omega_{m}=0$ (red curve), and $G_{a}/\omega_{m}=0.03$, $G_{m}/\omega_{m}=0$ (green curve).
To observe single-EIT behavior, we have considered the case when system is only coupled to the intra-cavity atomic mode or BEC and 
the coupling of moving-end mirror with intra-cavity optical mode is zero. The EIT behavior in cavity-optmechanics with coupled mirror have already been 
discussed in previous works like\cite{agarwal2010,Chang,Stefan2010,agarwal2013}. We consider that the optomechanical system is 
being operated in strong coupling regime
which means intra-cavity optical mode is strongly coupled to atomic mode of the system. When collective density excitation of 
atomic mode of cavity became resonant with intra-cavity optical mode, it cause strong coupling between atomic mode and system.
It is only possible when the strength of coupling between single atom and single photon of the cavity $g_0=10.9\times2\pi MHz$ is larger than both decay rate of 
the atomic excited state $\gamma_r=0.21\times2\pi kHz$ as well as intra-cavity field decay $\kappa=1.3\times2\pi kHz$ ($g_0>>\gamma_a,\kappa$)\cite{Esslinger,Stefan2010}.
We can also note from mathematical expression of atomic coupling that the strength of atomic mode coupling is directly proportional to the vacuum Rabi frequency 
($G_a\propto U_0=g^2_0/\Delta_a$).

In Fig.\ref{fig2a}, one can easily observe there are no signatures of EIT in absorption and dispersion spectra of output field (blue curves) when intra-cavity 
field is not coupled with mechanical mode and condensate mode ($G_{a}/\omega_{m}=0$, $G_{m}/\omega_{m}=0$). While in red curve, 
the single-EIT window appears in output probe field due to intra-cavity optical mode coupling with atomic mode (BEC) 
($G_{a}/\omega_{m}=0.02$), however, coupling of intra-cavity field with mechanical mode is kept zero $G_{m}/\omega_{m}=0$. Green 
curve shows the enhancement in single-EIT window in output probe field by increasing the coupling strength to $G_{a}/\omega_{m}=0.03$. 
These results clearly prove the existence of single-EIT window in output probe field when intra-cavity optical mode is only 
coupled to atomic mode of cavity-optomechanics.

\begin{figure*}[htp]
\centering
\includegraphics[width=18cm]{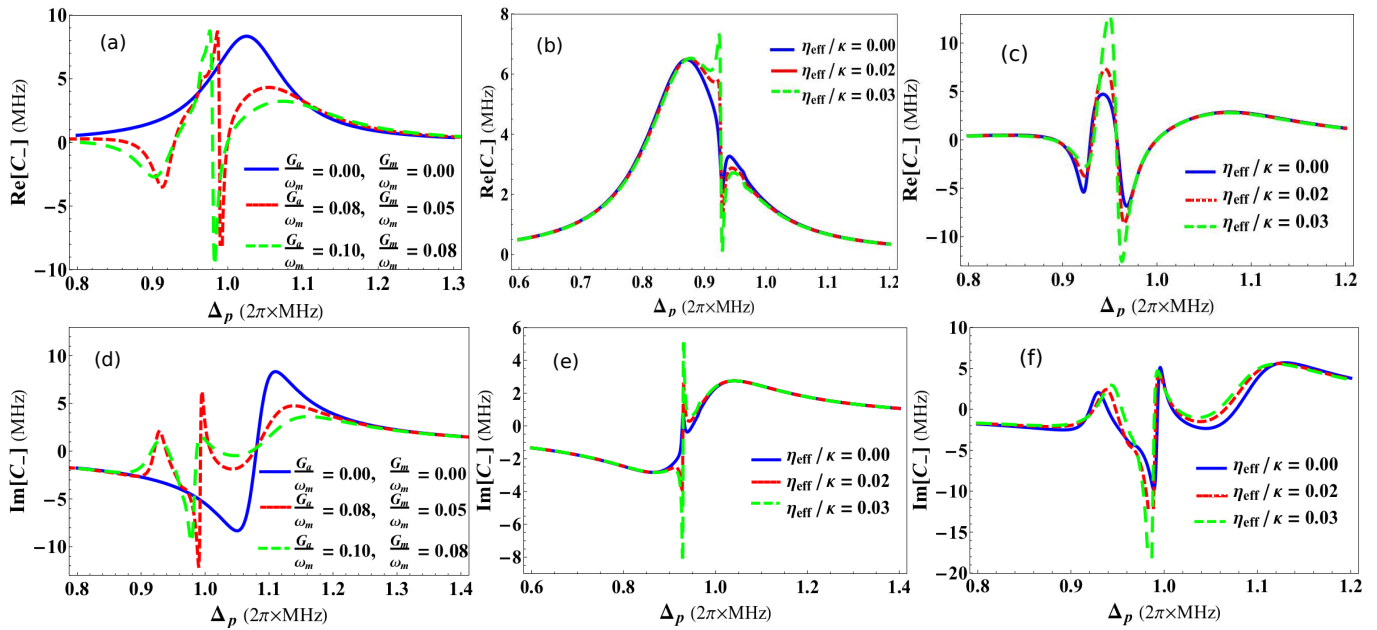}
\caption{(Color online) The real and imaginary quadratures of the output probe field $C_{-}$ are presented as a function of probe detuning 
$\Delta_{p}/\omega_{m}$ and transverse optical field $\eta_{eff}/\kappa$. (a) and (d) accommodate double-EIT behavior in 
real and imaginary 
quadratures, respectively, in the absence of transverse field $\eta_{eff}/\kappa=0$ and with various coupling strengths $G_{a}/\omega_{m}=0$, 
$G_{m}/\omega_{m}=0$ (blue curve), $G_{a}/\omega_{m}=0.08$, $G_{m}/\omega_{m}=0.05$ (red curve), and $G_{a}/\omega_{m}=0.1$, $G_{m}/\omega_{m}=0.08$ (green curve). 
(b) and (e) demonstrate single-EIT windows with coupling strengths $G_{a}/\omega_{m}=0.03$, $G_{m}/\omega_{m}=0$, as a function of 
transverse field strengths $\eta_{eff}/\kappa=0$ (blue curve), $\eta_{eff}/\kappa=0.02$ (red curve) and $\eta_{eff}/\kappa=0.03$ (green curve). 
Similarly, (c) and (f) show double-EIT behavior of output field spectra with coupling $G_{a}/\omega_{m}=0.1$, $G_{m}/\omega_{m}=0.08$, as a function of 
transverse field strengths $\eta_{eff}/\kappa=0$ (blue curve), $\eta_{eff}/\kappa=0.02$ (red curve) and $\eta_{eff}/\kappa=0.03$ (green curve). 
The remaining parameters are same as in Fig.\ref{fig2a}.}
\label{fig2}
\end{figure*}

Fig.\ref{fig2} describes the EIT behavior in the output field spectra of the optomechanical system in presence 
of probe detuning $\Delta_{p}/\omega_{m}$ and transverse optical field $\eta_{eff}/\kappa$. Fig.\ref{fig2}(a) and 
Fig.\ref{fig2}(d) represent absorption (real) and dispersion (imaginary) quadratures, respectively, of output probe field
in the absence of transverse field $\eta_{eff}/\kappa=0$ with various coupling strengths $G_{a}/\omega_{m}=0$, 
$G_{m}/\omega_{m}=0$ (blue curve), $G_{a}/\omega_{m}=0.08$, $G_{m}/\omega_{m}=0.05$ (red curve), and $G_{a}/\omega_{m}=0.1$, $G_{m}/\omega_{m}=0.08$ (green curve).
Results show that there are no signs of EIT-like behavior in absorption and dispersion spectra of output field (blue curves) when 
system is isolated form mechanical mode and condensate mode ($G_{a}/\omega_{m}=0$, $G_{m}/\omega_{m}=0$). On the other hand, 
in red curve, two EIT windows 
appear in output probe field because the optical mode of the system is now coupled to both mechanical mode (moving-end mirror) 
($G_{m}/\omega_{m}=0.05$) as well as to the condensate mode with coupling strength $G_{a}/\omega_{m}=0.08$. Such behavior is also 
known as double-EIT response of output field \cite{agarwal2013,Jadi2015}. When system is coupled to atomic mode and mechanical mode at the same time and optical 
mode of the system become resonant to both these modes, it gives rise 
to anti-Stokes scattering inside the system causing appearance of another EIT window in output field. 
Green curves in Fig.\ref{fig2}(a) and (d) demonstrate similar behavior double-EIT 
when the coupling strengths are increased to $G_{a}/\omega_{m}=0.1$, $G_{m}/\omega_{m}=0.08$. We observe that the quadratures of 
double-EIT behavior are increased by increasing coupling strengths.
Given results in Fig.\ref{fig2}(a) and (d) show such double-EIT behavior in output field when optomechanical system is coupled to 
moving-end mirror of the system and BEC trapped inside the system.

Fig.\ref{fig2}(b) and Fig\ref{fig2}(e) demonstrate single-EIT behavior in absorption and dispersion quadratures, respectively, of 
output probe field under the influence of transverse optical field $\eta_{eff}/\kappa$ when intra-cavity 
optical mode is coupled to condensate mode with coupling strength $G_{a}/\omega_{m}=0.03$ while the coupling of optical mode
with moving-end mirror is zero $G_{m}/\omega_{m}=0$. [Note: we cannot observe transverse optical field effects on single-EIT 
when intra-cavity optical degree of freedom is only coupled to the moving-end mirror, as shown in single-EIT results in 
previous works like \cite{agarwal2010,Chang,Stefan2010,agarwal2013}, because transverse optical field is only interacting with BEC trapped inside the 
cavity. Therefore, we only consider condensate mode coupling while studying effects of transverse field on single-EIT.] 
Blue curves show single-EIT windows in output probe field in the absence of transverse optical field $\eta_{eff}/\kappa=0$. 
On the other hand, red and green curves demonstrate the effects of transverse field strengths $\eta_{eff}/\kappa=0.02$ and 
$\eta_{eff}/\kappa=0.03$, respectively, on the single-EIT behavior. When transverse field photon interacts with atomic mode of 
the system, it gives rise to the total photon number $n$ inside the cavity by scattering transverse photons into the system, 
which leads to another nonlinear contribution to the anti-Stokes scatterings and enhance the EIT behavior in output field.
We can observe such effects of transverse field in the results that the strength of single-EIT is efficiently 
amplified by increasing the strength of transverse optical field. 

Similarly, Fig.\ref{fig2}(c) and Fig\ref{fig2}(f) represent double-EIT behavior in absorption and dispersion quadratures respectively, of 
output probe field as a function of transverse optical field $\eta_{eff}/\kappa$ when intra-cavity 
optical mode is coupled to both condensate mode with coupling strength $G_{a}/\omega_{m}=0.1$ and 
to the moving-end mirror is $G_{m}/\omega_{m}=0.08$. Blue curves in both these figures describe double-EIT behavior in the 
absence of transverse field $\eta_{eff}/\kappa=0$. Besides, red and green curves represent double-EIT with transverse field strengths 
$\eta_{eff}/\kappa=0.02$ and $\eta_{eff}/\kappa=0.03$, respectively. We can observe, like single-EIT results, double-EIT windows 
are enhanced by increasing the transverse optical coupling. Therefore, in accordance to these results, we can confidently 
state that by increasing transverse optical field coupling, we can enhance the phenomenon of electromagnetically induced 
transparency (EIT) in hybrid BEC-optomechanics.

\section{Tunable Fano resonances}\label{sec4}

\begin{figure}[tbp]
\includegraphics[width=7cm]{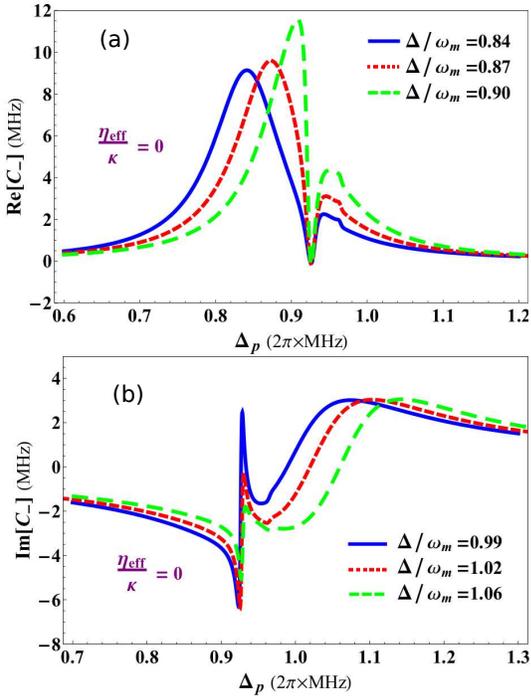}
\caption{(Color online) Single-Fano resonances are shown in the absorption (real) and dispersion (imaginary) profile of output probe field in the absence of 
transverse field coupling, as a function of normalized probe field detuning $\Delta_{p}/\omega_{m}$ and normalized 
effective detuning of the system $\Delta/\omega_{m}$ with coupling $G_{a}/\omega_{m}=0.03$ and $G_{a}/\omega_{m}=0$. 
(a) demonstrates absorption spectra 
of output field with different values of effective detuning $\Delta/\omega_{m}=0.84$ (blue curve), $\Delta/\omega_{m}=0.87$ (red curve) and 
$\Delta/\omega_{m}=0.90$ (green curve). Similarly, (b) shows dispersion behavior of output field with effective detuning 
$\Delta/\omega_{m}=0.99$ (blue curve), $\Delta/\omega_{m}=1.02$ (red curve) and 
$\Delta/\omega_{m}=1.06$ (green curve). All other parameters are same as in Fig.\ref{fig2a}.}
\label{fig3}
\end{figure}

\begin{figure}[htp]
\includegraphics[width=7cm]{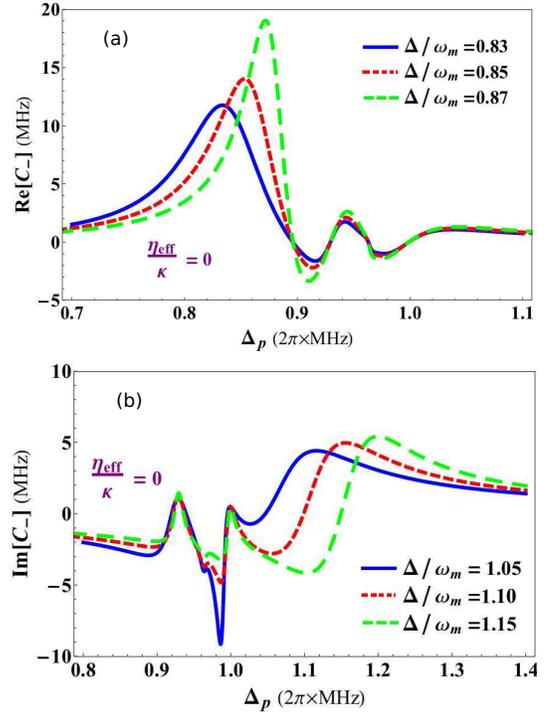}
\caption{(Color online) Double-Fano resonances are presented in the absorption (real) and dispersion (imaginary) profile of output probe field in the absence of 
transverse field coupling, as a function of normalized probe field detuning $\Delta_{p}/\omega_{m}$ and normalized 
effective detuning of the system $\Delta/\omega_{m}$ with coupling $G_{a}/\omega_{m}=0.1$ and $G_{a}/\omega_{m}=0.08$.
(a) shows real quadrature 
of output field under different effective detuning strengths $\Delta/\omega_{m}=0.83$ (blue curve), $\Delta/\omega_{m}=0.85$ (red curve) and 
$\Delta/\omega_{m}=0.97$ (green curve). On the other hand, (b) presents imaginary quadrature of output field with effective detuning values 
$\Delta/\omega_{m}=1.05$ (blue curve), $\Delta/\omega_{m}=1.1$ (red curve) and 
$\Delta/\omega_{m}=1.15$ (green curve). Remaining parameters are same as in Fig.\ref{fig2a}.}
\label{fig4}
\end{figure}

The formation of Fano resonance in the output optical mode of hybrid optomechanical system is a fascinating phenomenon caused 
by quantum mechanical interaction between different degrees of freedom of the system \cite{agarwal2013,Jadi2015}. The constructive and destructive 
quantum interferences among narrow discrete intra-cavity optical resonances are the foundation of Fano resonances in output 
of such complex systems. The transverse field effects on EIT presented in Fig.\ref{fig2a} and Fig.\ref{fig2} are similar to the single and 
double-Fano resonances but tuned by transverse optical field.  
We conventionally observe Fano line shapes in EIT windows by tunning effective detuning of the system. The variation in 
effective detuning of the system brings change to the anti-Stokes scatterings which causes the shift in EIT window. 
In following, we demonstrate Fano behavior of system output field with respect to different parameters.

Fig.\ref{fig3} shows single-Fano resonances in the absorption (real) and dispersion (imaginary) profile of output probe field in the absence of 
transverse field coupling $\eta_{eff}/\kappa=0$, as a function of normalized probe field detuning $\Delta_{p}/\omega_{m}$ and normalized 
effective detuning of the system $\Delta/\omega_{m}$. The coupling of intra-cavity optical mode with atomic mode 
is $G_{a}/\omega_{m}=0.03$ while, the coupling of optical mode with mechanical mode is kept zero ($G_{m}/\omega_{m}=0$). 
which means, optomechanical system is only coupled to the condensate mode trapped inside the cavity.
Fig.\ref{fig3}(a) and Fig.\ref{fig3}(b) describe absorption and dispersion profile,respectively, in output probe field 
as a function of normalized probe detuning. Blue curve in absorption shows Fano line with effective system 
detuning $\Delta/\omega_{m}=0.84$.
While, red and green curves in real quadrature represent fano behavior under influence of effective detuning $\Delta/\omega_{m}=0.87$ and $\Delta/\omega_{m}=0.9$, 
respectively. Similarly, blue curve in dispersion profile shows the existence of Fano resonance with effective system 
detuning $\Delta/\omega_{m}=0.99$.
Besides, red and green curves in imaginary quadrature of output field represent fano behavior under influence of effective detuning 
$\Delta/\omega_{m}=1.02$ and $\Delta/\omega_{m}=1.06$, respectively.
We can observe, each curve with different height follow a same dip in absorption and dispersion response which causes 
the formation of resonance in out-going optical mode. By analyzing these curves, one can predict the formation of Fano 
resonance in the output field.

\begin{figure}[tp]
\includegraphics[width=7cm]{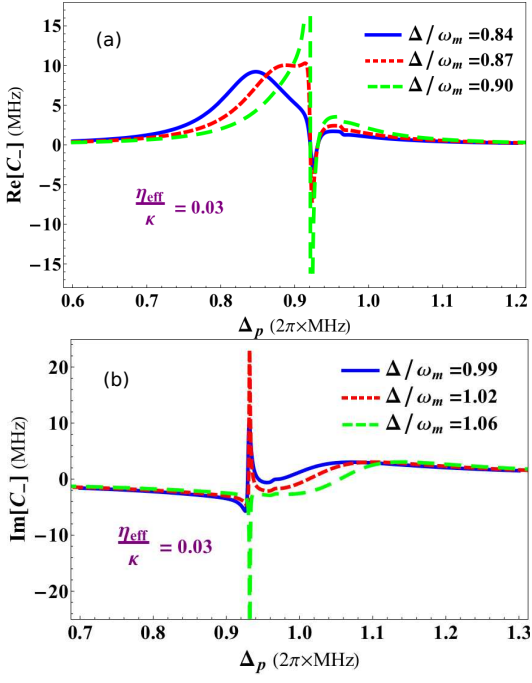}
\caption{(Color online) The effects of transverse optical field, with coupling $\eta_{eff}/\kappa=0.03$, on single-Fano resonances are presented in the absorption (real) 
and dispersion (imaginary) profiles of output probe field as a function of normalized probe field detuning $\Delta_{p}/\omega_{m}$ and 
under the influence of normalized effective detuning of the optomechanical system $\Delta/\omega_{m}$. 
The coupling strengths are same as in Fig.\ref{fig3}.
(a) shows absorption quadrature 
of output optical mode with different detuning strengths $\Delta/\omega_{m}=0.84$ (blue curve), $\Delta/\omega_{m}=0.87$ (red curve) and 
$\Delta/\omega_{m}=0.9$ (green curve). While, (b) accounts for imaginary quadrature of output field having effective detuning strengths 
$\Delta/\omega_{m}=0.99$ (blue curve), $\Delta/\omega_{m}=1.02$ (red curve) and 
$\Delta/\omega_{m}=1.06$ (green curve). Remaining parameters are same as in Fig.\ref{fig2a}.}
\label{fig5}
\end{figure}

\begin{figure}[htp]
\includegraphics[width=7cm]{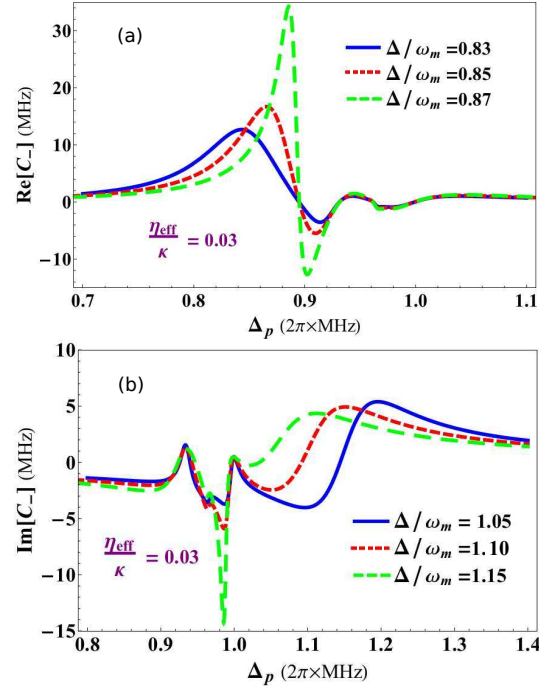}
\caption{(Color online) The behavior of double-Fano resonances are demonstrated in the absorption (real) 
and dispersion (imaginary) quadratures of output probe field as a function of normalized probe field detuning $\Delta_{p}/\omega_{m}$ and 
effective detuning $\Delta/\omega_{m}$
under the notable effects of transverse optical field having strength $\eta_{eff}/\kappa=0.03$. 
The system coupling strengths are same as in Fig.\ref{fig4}. 
(a) shows absorption profile 
of output field under effective detuning strengths $\Delta/\omega_{m}=0.83$ (blue curve), $\Delta/\omega_{m}=0.85$ (red curve) and 
$\Delta/\omega_{m}=0.97$ (green curve). On the other hand, (b) presents dispersion quadrature of output field with effective detuning values 
$\Delta/\omega_{m}=1.05$ (blue curve), $\Delta/\omega_{m}=1.1$ (red curve) and 
$\Delta/\omega_{m}=1.15$ (green curve). All other parameters are same as in Fig.\ref{fig2a}.}
\label{fig6}
\end{figure}
We further investigate the existence of double-Fano resonances in output probe field by introducing another coupling in the 
optomechanical system and modifing effective detuning \cite{Jadi2015}. As the phenomenon of EIT is very sensitive to the coupling with different degrees 
of freedom in the system. Therefore by introducing another coupling, we can convert single-Fano resonance to double-Fano 
resonance. Fig.\ref{fig4} shows such double-Fano resonances in the absorption and dispersion profile of output probe field in the absence of 
transverse field coupling $\eta_{eff}/\kappa=0$ and as a function of normalized probe field detuning $\Delta_{p}/\omega_{m}$ and normalized 
effective detuning of the system $\Delta/\omega_{m}$. The coupling of intra-cavity optical mode with moving-end mirror 
is $G_{a}/\omega_{m}=0.1$ and the coupling of optical mode with condensate mode is $G_{a}/\omega_{m}=0.08$.  
Fig.\ref{fig4}(a) describes absorption and Fig.\ref{fig4}(b) describes dispersion profile in output probe field 
as a function of normalized probe detuning. Blue curves, in Fig.\ref{fig4}(a) and Fig.\ref{fig4}(b), show the double-Fano 
line with effective detuning values $\Delta/\omega_{m}=0.85$ and $\Delta/\omega_{m}=1.1$, respectively.
Similarly, red curves, in absorption and dispersion, accommodate the double-Fano response under the influence of effective detuning 
$\Delta/\omega_{m}=0.94$ and $\Delta/\omega_{m}=0.96$, respectively and green curves accounts for the influence of 
effective detuning $\Delta/\omega_{m}=0.87$ and $\Delta/\omega_{m}=1.15$, respectively. By analyzing these results, we come to know the 
existence of single-Fano resonances as well as double-Fano resonances in the output probe field of the system \cite{agarwal2013,Jadi2015}. 

In previous Fano resonance results, we have ignored the effects of transverse optical field coupling. But it will be very 
important to keep these effects and analyze the behavior of Fano resonances. Fig.\ref{fig5} illustrates such effects on single-Fano 
resonances emerging in output field spectra in the presence of transverse field coupling $\eta_eff/\kappa=0.03$. 
Fig.\ref{fig5}(a) shows real quadrature of out-going mode where, blue, red and green 
curves corresponds to the effective detuning strengths $\Delta/\omega_{m}=0.85$, $\Delta/\omega_{m}=0.87$ and $\Delta/\omega_{m}=0.9$, 
respectively. On the other hand, Fig.\ref{fig5}(b) represents imaginary quadrature of output field where, blue, red and green 
curves correspond to the influence of system detuning strengths $\Delta/\omega_{m}=0.99$, $\Delta/\omega_{m}=1.02$ and $\Delta/\omega_{m}=1.06$, 
respectively. One can observe, how quadratures of singe-Fano lines are increased due to the presence of transverse field. Transverse 
optical field causes scattering of photon inside the cavity which gives rise to intra-cavity photon number and this nonlinear factor 
brings modification to the out-going optical mode of cavity. It is understood that if we further increase the strength of transverse 
coupling, it will definitely modify Fano behavior in output field.

Fig.\ref{fig6} shows similar 
behavior for double-Fano resonances in real and imaginary profile of out-going probe field under the influence of transverse 
field strength. Fig.\ref{fig6}(a) shows absorption and Fig.\ref{fig6}(b) shows dispersion behavior at transverse optical 
field strength $\eta_{eff}/\kappa=0.03$. Blue curves, in Fig.\ref{fig6}(a) and Fig.\ref{fig6}(b), show the effects of $\eta_{eff}/\kappa$ 
on double-Fano 
curves appearing in out-going mode with effective detuning values $\Delta/\omega_{m}=0.85$ and $\Delta/\omega_{m}=1.1$, respectively.
While, red curves, in absorption and dispersion quadratures, illustrates the effects on double-Fano response with effective detuning 
$\Delta/\omega_{m}=0.94$ and $\Delta/\omega_{m}=0.96$, respectively and green curves accounts for the similar response 
double-Fano resonance under the influence of 
effective detuning $\Delta/\omega_{m}=0.87$ and $\Delta/\omega_{m}=1.15$, respectively.
By comparing results of Fig.\ref{fig3} and Fig.\ref{fig4} with Fig.\ref{fig5} and Fig.\ref{fig6}, we can easily note the effects of transverse optical field 
on the double-Fano resonance of the optomechanical system. The absorption and dispersion quadratures of single-Fano as well 
as double-Fano resonances are notably modified by 
increasing transverse optical field strength which we can observe in presented results.

\section{Conclusion}\label{sec5}

In conclusion, we discuss the controllability of electromagnetically induced transparency (EIT) and Fano Resonances in hybrid 
BEC-optomechanical system which is composed of cigar-shaped Bose-Einstein condensate (BEC) trapped inside high-finesse 
Fabry-P\'{e}rot cavity driven by a single mode optical field along the cavity axis and a transverse pump field. 
As, the transverse optical field directly interacts with condensate mode which causes the scattering of transverse photon inside the cavity, so, 
by varying transverse field, we can modify the dynamics of system. We have shown 
the controlled behavior of electromagnetically induced transparency (EIT) in output probe field by using transverse field. 
We discuss existence of single-EIT window in output field of cavity-optomechanics in the absence of moving-end mirror, which means 
intra-cavity optical mode was only coupled to atomic mode (BEC) of the system.
The single-EIT as well as double-EIT windows in output probe laser field are efficiently amplified by increasing 
the strength of transverse optical field. Furthermore, the single and double-Fano resonances are discussed in out-going 
probe field of the system. The influence of transverse optical field is also been studied on Fano resonances of the system. 
Moreover, we have suggested a certain set of experimental parameters to observe these phenomena in laboratory.

In future, we will apply this method of control to discuss the nonlinear dynamics of hybrid system, especially to explore 
dynamics of interacting Bose-Einstein condensate in such complex systems. Besides, we intend to extend this method to study the controllability of novel
phenomenon like entanglement. Additional goals include the study of effects of 
spin-orbit coupling \cite{spinorbit,Dong14} using magnetic field in hybrid BEC-optomechanical systems.

\begin{acknowledgments}
 This work was supported by the NKBRSFC under grants Nos. 2011CB921502, 2012CB821305, NSFC under grants 
Nos. 61227902, 61378017, 11434015, SKLQOQOD under grants No. KF201403, SPRPCAS under grants No. XDB01020300.
We strongly acknowledge financial support from CAS-TWAS President's PhD fellowship programme (2014).
\end{acknowledgments}

\end{document}